\documentclass[twocolumn,showpacs,prb,amsmath,amssymb]{revtex4-1}
\usepackage{graphicx}% Include figure files
\usepackage{booktabs}
\usepackage{multirow}
\usepackage{amssymb}
\usepackage{longtable}

\begin{document}
\title{1.5 GHz Pulse Generation From a Monolithic Waveguide Laser With a Graphene-Film Saturable Output Coupler}
\author{R. Mary$^{1}$, S.J. Beecher$^{1}$, G. Brown$^{1}$, F. Torrisi$^2$, S. Milana$^2$, D.Popa$^2$, T. Hasan$^2$, Z. Sun$^2$, E. Lidorikis$^3$, S. Ohara$^4$, A.C. Ferrari$^2$ and A.K. Kar$^{1}$}

\affiliation{$^1$SUPA, Institute of Photonics and Quantum Sciences, School of Engineering and Physical Sciences, Heriot-Watt University, EH14 4AS, UK\\
$^2$Department of Engineering, University of Cambridge, Cambridge, CB3 0FA, UK\\
$^3$Department of Materials Science and Engineering, University of Ioannina, Ioannina, Greece\\
$^4$Asahi Glass Co., Ltd. Research Center, 1150, Hazawa-cho, Kanagawa-ku, Yokohama, Kanagawa 221-8755, Japan\\}
\begin{abstract}
We fabricate a saturable absorber mirror by coating a graphene film on an output coupler mirror. This is then used to obtain Q-switched mode-locking from a diode pumped linear cavity waveguide laser inscribed in Ytterbium-doped Bismuthate Glass, with high slope and optical conversion efficiencies. The laser produces mode-locked pulses at$\sim$1039nm, with 1.5GHz repetition rate at an average 202mW output power. This performance is due to the combination of the graphene saturable absorber with the high quality laser glass.
\end{abstract}
\maketitle
Carbon nanotubes (CNTs) and graphene have emerged as promising saturable absorbers (SA) for a variety of applications\cite{Hasan_am_2009,Bonaccorso_np_2010,Sun_pe_2012}, opening a new phase in the development of passively Q-switched\cite{Popa_apl_2011} and mode-locked lasers\cite{Popa_apl_2012,Sun_an_2010,Popa_apl_2010,Sun_nr_2010,Zhang_oe_2012,Ma_ol_2012,Zhao_ol_2010,Wang_nn_2008,Valle_apl_2006,Beecher_apl_2010,Lagatsky_2012}. While the predominantly used semiconductor saturable absorber mirrors (SESAMs) are limited by their narrow wavelength range\cite{Okhotnikov_njp_2004}, and complex fabrication\cite{Keller_po_2004}, CNTs and graphene have simple, cost-effective production and integration\cite{Hasan_am_2009,Bonaccorso_np_2010,Popa_apl_2010,Sun_pe_2012,Popa_apl_2011,Popa_apl_2012,Sun_an_2010,Sun_nr_2010,Zhang_oe_2012,Ma_ol_2012,Zhao_ol_2010,Wang_nn_2008,Valle_apl_2006,Beecher_apl_2010,Lagatsky_2012,bonamt}. Broadband operation is achieved with CNTs by combining tubes of different diameters\cite{Wang_nn_2008}. However, for a particular wavelength, only CNTs in resonance are used, the rest contributing insertion losses\cite{Sun_an_2010}. Graphene has an inherent ultra-wide spectral range due to the linear dispersion of the Dirac electrons\cite{Hasan_am_2009,Bonaccorso_np_2010,Sun_pe_2012,Popa_apl_2011,Sun_an_2010,Sun_nr_2010}. This, along with the ultrafast recovery time\cite{Brida_2012} and low saturation fluence\cite{Popa_apl_2010,Sun_an_2010}, makes it an excellent SA\cite{Hasan_am_2009,Sun_an_2010,Sun_nr_2010,Zhang_oe_2012,Zhao_ol_2010,Popa_apl_2010}.

Saturable absorption can also lead to a regime of Q-switched mode-locking (QML)\cite{Honninger_josab_1999}, where the laser output consists of mode-locked pulses within a Q-switched envelope\cite{Honninger_josab_1999}. This arises due to Q-switching instabilities in the cavity\cite{Honninger_josab_1999}, typically due to long (i.e.$>$1$\mu$s) upper state lifetimes of the gain media in solid state lasers\cite{Honninger_josab_1999}. These lasers are useful for applications where the pulse energy stored within the Q-switched envelope\cite{Honninger_josab_1999} is valuable, such as nonlinear frequency conversion\cite{Boyd_no}, medical applications\cite{Chung_jb_2009}, and micromachining\cite{Nolte_josab_1997}. With the emerging trend in miniaturization of optical devices based on on-chip integration, the development of ultrafast lasers requires a complementary balance between device compactness and performance\cite{Valle_apl_2006,Beecher_apl_2010}. Ultra-compact high repetition rate ($>$1GHz) lasers are very useful for applications such as nonlinear microscopy\cite{Mertz_con_2004}, frequency combs\cite{Diddams_josab_2010} and spectroscopy\cite{Moreno_josab_2011}. The ease of SA integration into a compact cavity plays an important role\cite{Valle_apl_2006,Beecher_apl_2010}. Lasers employing a waveguide cavity allow device compactness, meanwhile emulating the advantages of fiber lasers, such as high beam quality\cite{Okhotnikov_njp_2004} and efficient heat dissipation\cite{Valle_apl_2006,Beecher_apl_2010,Okhotnikov_njp_2004}. Of the numerous methods for waveguide fabrication, a simple yet reliable technology is ultrafast laser inscription (ULI)\cite{Valle_joa_2009}, which utilizes$\sim$100fs focused pulses to induce permanent modifications within a substrate\cite{Valle_joa_2009}. Mode-locked ULI waveguide lasers have been demonstrated using CNT-SAs\cite{Valle_apl_2006,Beecher_apl_2010}. However, the fiber ring cavity\cite{Valle_apl_2006,Beecher_apl_2010} did not allow its miniaturization, thus high pulse rates.

Here we report pulse generation in a compact, ULI inscribed waveguide laser in Ytterbium-doped Bismuthate Glass (Yb:BG), by using a graphene film (GF) transferred to an output coupler (OC) mirror as a SA. We get mode-locked pulses with 1.5GHz repetition rate and 202mW average output power, with a 48\% slope efficiency (i.e. rate of output to pump power in excess of the lasing threshold\cite{Grivas_pqe_2011}) and 38\% optical-to-optical conversion efficiency (i.e. rate of output to pump power\cite{Grivas_pqe_2011}). The slope efficiency is high compared with that typical of monolithic pulsed waveguide lasers (e.g. 27\%)\cite{Choudhary_ol_2012}.

A variety of approaches have been used to make graphene-based SAs. For example, graphene-polymer composites, fabricated from dispersions produced by liquid phase exfoliation (LPE) of graphite\cite{bonamt}, have been used to mode-lock fiber lasers at 1.5\cite{Hasan_am_2009,Popa_apl_2010,Sun_nr_2010} and 2$\mu$m\cite{Zhang_oe_2012}. Films grown by chemical vapour deposition (CVD) with 1 layer\cite{Lagatsky_2012}, 1-2 layers\cite{Ma_ol_2012}, and non-uniform multi-layers\cite{Zhao_ol_2010}, have been used to mode-lock solid-state lasers at 2$\mu$m\cite{Ma_ol_2012,Lagatsky_2012} and fiber lasers at 1$\mu$m\cite{Zhao_ol_2010}. Ref.\onlinecite{Martinez_apl_2011} used flakes produced by micromechanical cleavage of graphite, with 4-40 layers, for mode-locking of fiber lasers at 1.5$\mu$m. LPE graphene-polymer composites\cite{Popa_apl_2011}, and flakes (10-40 layers) grown by carbon segregation on SiC\cite{Yu_an_2010}, have been used for Q-switching of fiber lasers at 1.5$\mu$m and solid-state lasers at 1$\mu$m, respectively. Graphene oxide (GO) was also used as SA, either as a film in solid-state lasers at 2$\mu$m\cite{Liu_lpl_2012}, or as composite in fiber lasers at 1.5$\mu$m\cite{Bonaccorso_np_2010}. However, GO is an insulating material with many defects and gap states\cite{Mattevi_afm_2009}, and may not offer the wideband tunability of graphene. Carbon segregation and CVD require high substrate temperatures\cite{bonamt,Ma_ol_2012,Lagatsky_2012,Liu_lpl_2012}, followed by transfer to the target substrate\cite{Ma_ol_2012,Zhao_ol_2010,Lagatsky_2012}. Micromechanically cleaved graphene has high structural and electronic quality\cite{Bonaccorso_np_2010}, but is limited in terms of yield, thus impractical for large-scale applications\cite{bonamt}. LPE has the advantage of scalability, room temperature processing and high yield, and does not require any growth substrate\cite{bonamt}. Dispersions produced by LPE can easily be embedded into polymers composites and integrated into various systems\cite{Bonaccorso_np_2010,bonamt}. 

Here we adopt a novel approach and use LPE graphene in a polymer-free film. This makes it suitable for high-power applications and device miniaturization. The GF-SA is prepared as follows. Graphite flakes (Sigma Aldrich) undergo LPE\cite{Hernandez_nn_2008} and are dispersed in deionised water with sodium deoxycholate, as for Refs.\onlinecite{Sun_an_2010,Zhang_oe_2012}. High Resolution Transmission Electron Microscopy (HRTEM), optical and Raman Spectroscopy are then used to characterize the dispersions. HRTEM shows that the sample consists of$\sim$26\% single-,$\sim$22\% bi- and$\sim$18\% tri-layers\cite{Zhang_oe_2012,Hasan_pssb_2010}, with$\sim$1$\mu$m average size. The dispersion then undergoes vacuum filtration via 25nm pore-size filters. This blocks the flakes, while allowing water to pass through, resulting in a GF. This is then placed on an OC mirror, to be used in the laser, and on a quartz plate, for optical characterization, by applying pressure and heat ($\sim$80$^\circ$C, to improve adhesion) for two hours, followed by dissolution of the filter in acetone. The film is$\sim$45nm thick, as determined by profilometry. The GF density is$\sim$0.72g/cm$^3$, derived by measuring with a microbalance the filter weight before and after the GF deposition. This is$\sim$3 times smaller than the density of bulk graphite.
\begin{figure}
\centerline{\includegraphics[width=90mm]{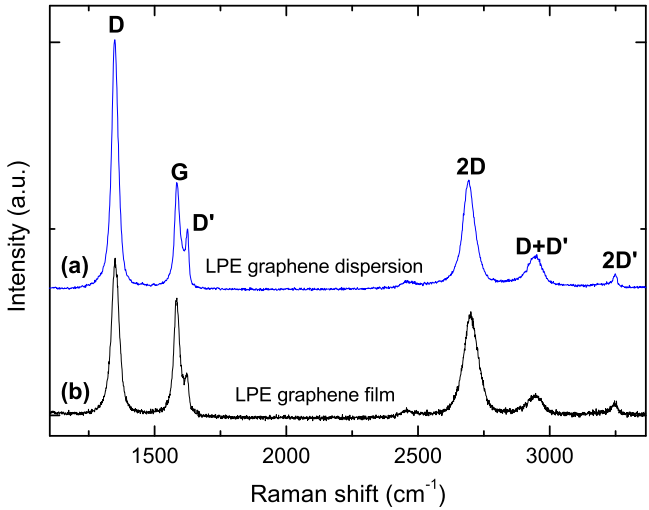}}
\caption{\label{raman} Raman spectra measured at 514nm of (a) graphene dispersion in SDC-Water and (b) graphene film.}
\end{figure}

Raman spectra are acquired at 457, 514, 633nm using a Renishaw InVia micro-Raman spectrometer. Fig.\ref{raman}(a) plots a typical Raman spectrum of graphene flakes in the dispersion. Besides the G and 2D peaks, significant D and D' bands are also present\cite{Ferrari_prl_2006,Ferrari_prb_2000}. We assign the D and D' peaks to the sub-micrometer edges of our flakes\cite{Casiraghi_nl_2009}, rather than to a large amount of disorder within the flakes. This is supported by the G peak dispersion, Disp(G)=0.02cm$^{-1}$/nm, much lower than in disordered carbons\cite{Ferrari_prb_2001}. Fig.\ref{raman}(b) plots the GF Raman spectrum at 514nm. Similar to the individual flakes discussed above, Disp(G) is 0.02 cm$^{-1}$/nm\cite{Ferrari_prb_2001}. The 2D peak is still single Lorentzian, but$\sim$24cm$^{-1}$ larger than for the individual flakes. Thus, even if the flakes are multi-layers, they are electronically decoupled and, to a first approximation, behave as a collection of single layers\cite{Hasan_pssb_2010,Latil_prb_2007}. The ratio of the 2D and G integrated areas, A(2D)/A(G), is at most$\sim$2, thus we estimate a doping$\sim$1.3x10$^{13}$cm$^{-2}$ [\onlinecite{Basko_prb_2009}], i.e. a Fermi level shift$\sim$4-500meV\cite{Das_nn_2008,Basko_prb_2009}.
\begin{figure}
\centerline{\includegraphics[width=90mm]{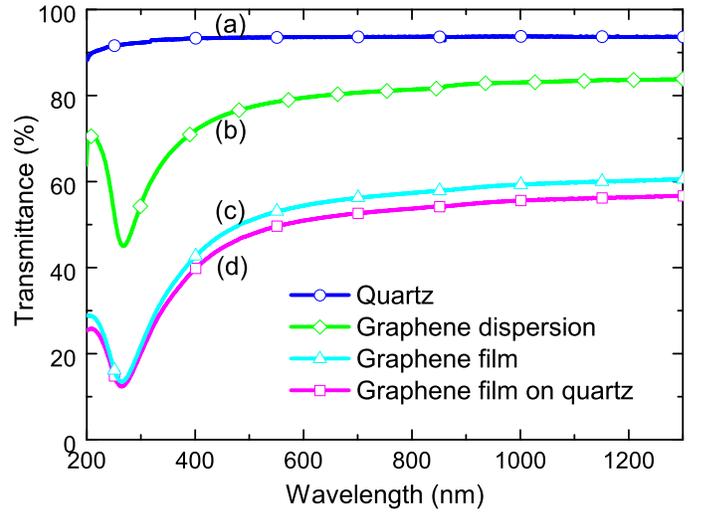}}
\caption{\label{trans} Transmittance of (a) quartz, (b) graphene dispersion, (c) graphene-film; (d) graphene-film on quartz.}
\end{figure}

Fig.\ref{trans}(b) plots the transmittance of the graphene dispersion (diluted to 10\% to avoid scattering losses at higher concentrations). Using T = e$^{-\alpha lc}$ where l[m] is the light path length, c[gL$^{-1}$] is the concentration of dispersed graphitic material, and $\alpha$[Lg$^{-1}$m$^{-1}$] is the absorption coefficient, with $\alpha$ $\sim$1390Lg$^{-1}$m$^{-1}$ at 660nm\cite{Lotya_jacs_2009}, we derive c$\sim$0.18gL$^{-1}$. The peak$\sim$266nm is a signature of the van Hove singularity in the graphene density of states\cite{Kravets_prb_2010}. Fig.\ref{trans}(a,c,d) plot the transmittance of quartz, pure GF and GF on quartz. The transmittance and reflectance at 1039nm (the laser wavelength) are$\sim$59\% and$\sim$11\% respectively. To estimate the number of graphene layers from these measurements we use the recurrent matrix method, including the correction to the graphene optical conductivity induced by doping\cite{Mak_prl_2008}. While pristine graphene absorbs 2.3\% per layer, doping, and consequent Pauli blocking, can significantly decrease this\cite{Sun_an_2010,Li_np_2008}. By comparing our calculations with the data at 1039nm we estimate that our GF consists of$\sim$40 layers. Taking into account that its density is$\sim$1/3 of graphite, this number of layers correspond to an overall thickness$\sim$40nm, in good agreement with that measured by profilometry. We note that a 40nm thick undoped and compact GF would absorb 100\% of the incident light and be near impossible to saturate, thus the low density and doping of our film are essential for the SA to work.
\begin{figure}
\centerline{\includegraphics[width=90mm]{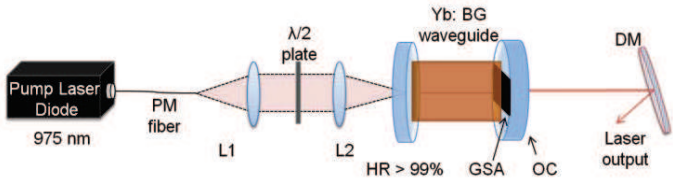}}
\caption{\label{setup} Laser schematic. L1 and L2: coupling lenses; PM: polarization maintaining fiber; GSA: Graphene saturable absorber; OC: output coupler; DM: Dichroic mirror.}
\end{figure}

Fig.\ref{setup} is the schematic of our cavity.  We use a 50mm Yb:BG substrate with 1.6x10$^{26}$m$^{-3}$Yb$^{3+}$ dopants and 2.03 refractive index as gain medium. The waveguide is inscribed by focusing the pulses, through a 0.4NA lens, 200$\mu$m below the substrate surface, by a master oscillator power amplifier fiber laser (IMRA FCPA $\mu$-Jewel D400) delivering 350fs pulses at 1047nm and 1MHz repetition rate.  An automated x-y-z stage translates the sample, thus extending the positive index change at the laser focus to form a waveguide. Low insertion loss waveguides with symmetric cross-sections are realised using a multi-scan technique\cite{Nasu_ol_2005}, inscribed with pulse energies$\sim$50nJ. Previously, highly efficient continuous wave lasing was demonstrated from these, with top slope efficiency$\sim$79\%\cite{Mary_ol_2012}.
\begin{figure}
\centerline{\includegraphics[width=90mm]{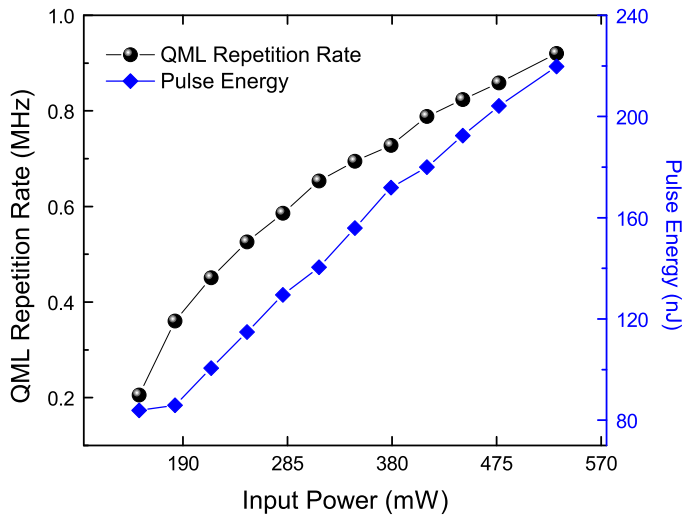}}
\caption{\label{rr} Repetition rate, and pulse energy within a single Q-switched envelope as a function of input pump power.}
\end{figure}

The pump source is a polarisation-maintaining fiber-coupled diode laser at 976nm, with 530mW maximum pump power, and an angle cleaved fiber to avoid back reflections. Two identical lenses with a 6.2nm focal length couple the pump light efficiently into the waveguide, and a half-wave plate varies the pump polarization. A dichroic mirror with 99\% reflection from 1010-1200nm and $<$2\% at the pump is the pump mirror. The mirrors are butt-coupled to either waveguide end using an index matching gel, which also reduces the parasitic Fresnel reflections at the interfaces\cite{Bass_foh}. A dichroic mirror separates the QML output from the residual pump light.

\begin{figure}
\centerline{\includegraphics[width=90mm]{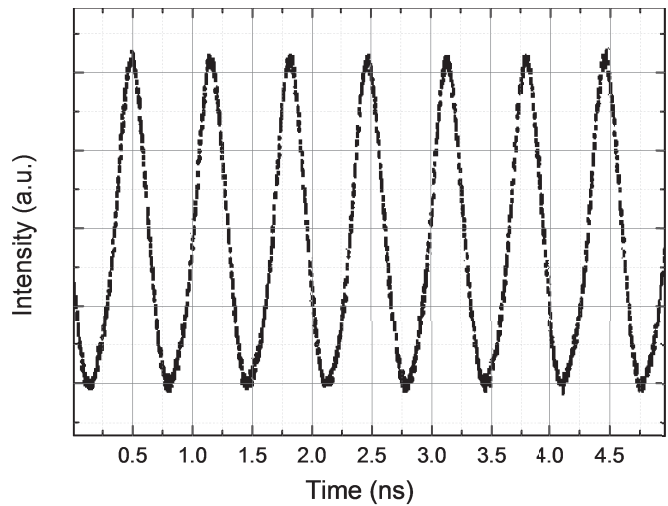}}
\caption{\label{osc} Mode-locked pulse train.}
\end{figure}

\begin{figure}
\centerline{\includegraphics[width=90mm]{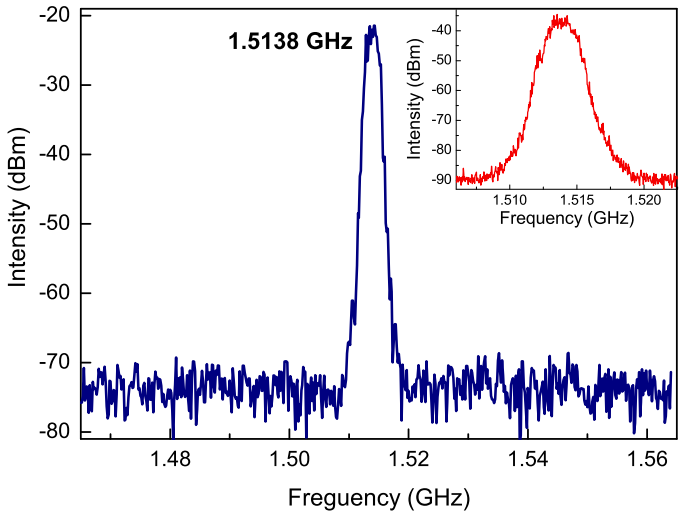}}
\caption{\label{rf} RF Spectrum measured at the maximum pump.}
\end{figure}

The laser operation initiates abruptly at a threshold pump power of 100mW in a self-starting QML regime. The cavity is optimized by adjusting the pump coupling efficiency, pump beam polarization, and GF-SA position. The mode area on the GF-SA is dictated by that of the waveguide. Using a fast photodiode and a wide-bandwidth oscilloscope, the initial QML repetition rate is measured as 200kHz, with 17mW average output power. Mode-locked pulses at a fundamental repetition rate of 1.514GHz, corresponding to the free spectral range of the cavity, are measured within the Q-switched envelope. Fig.\ref{rr} shows the QML pulse repetition rate and energy evolution within a single Q-switched envelope. As the launch pump power is increased, the period between the Q-switched pulses reduces, indicating a tendency towards CW mode-locking. At the highest available pump power of 530mW, the Q-switching modulation has a frequency of 0.95MHz, and an average output power of 202mW, corresponding to a pulse energy of 220nJ. This is distributed along the mode-locked pulses existing within the Q-switch envelope. Fig.\ref{osc} shows a constant mode-locked pulse train behaviour measured on a timescale of 500ps/div. Mode-locking at the fundamental repetition rate is also verified by measuring the rf spectra with a Rigol 1030 spectrum analyzer (Fig.\ref{rf}). The$\sim$4.2MHz spectral width indicates no pure CW mode-locking\cite{Linde_apb_1986}, as further shown in the inset of Fig.\ref{osc}.
\begin{figure}
\centerline{\includegraphics[width=90mm]{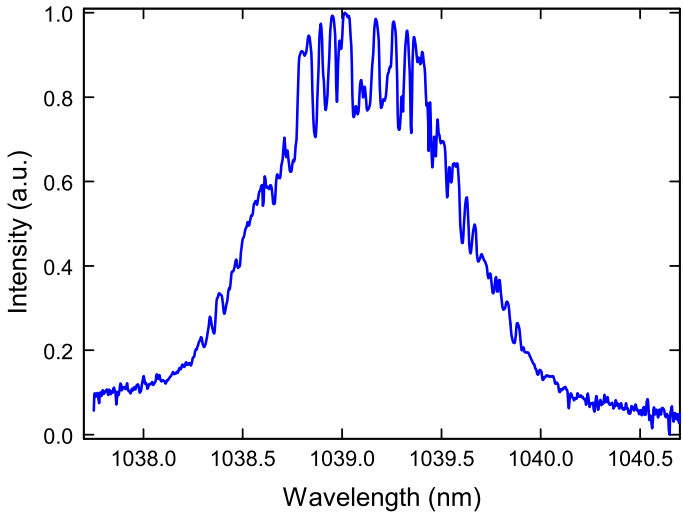}}
\caption{\label{spectrum} Optical Spectrum.}
\end{figure}
\begin{figure}
\centerline{\includegraphics[width=80mm]{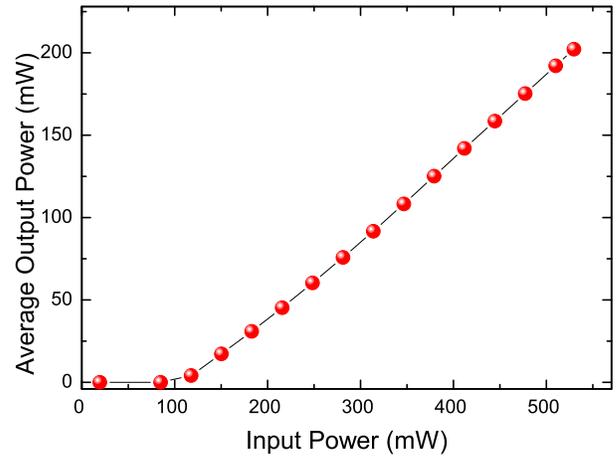}}
\caption{\label{in-out} Output power. Slope efficiency$\sim$48\% (38\% optical-to-optical efficiency). The highest output power is 202mW.}
\end{figure}

The optical spectrum is given in Fig.\ref{spectrum}. The spectral bandwidth, corresponding to a pump power of 530mW, is 1.1nm. With increasing pump, the spectral peak migrates slightly to longer wavelengths. For the maximum input pump power of 530mW, we have an average output power of 202mW. The average output power dependence on the pump is given in Fig.8. The QML waveguide laser has a high slope efficiency of 48\%, and a 38\% overall optical-to-optical conversion efficiency. Stable QML pulses are observed over$\sim$24 hours, establishing the good quality of the GF-SA.

In conclusion, a monolithic waveguide laser with stable and efficient Q-switched mode-locking was demonstrated using a transferred graphene film to an output coupler. This is a robust, reliable, practical passive mode-locking element, with easy integration in a waveguide cavity.

We acknowledge funding from ERC Grant NANOPOTS, EPSRC Grant EP/G030480/1, a Royal Society Wolfson Research Merit Award, The Royal Academy of Engineering, Emmanuel College and King's College, Cambridge.

\end{document}